\documentclass[conference]{IEEEtran}
\IEEEoverridecommandlockouts
\usepackage{graphicx}

\DeclareGraphicsExtensions{.pdf,.jpeg,.png}

\usepackage{balance}
\usepackage{url}
\usepackage{xcolor}
\usepackage{threeparttable}
\usepackage{float}
\usepackage{booktabs}
\usepackage{tabularx}         
\usepackage[tight]{subfigure}
\usepackage{caption}
\usepackage[binary-units]{siunitx}
\usepackage{acronym}
\usepackage{pgfplots}
\usepackage{multirow}
\usepackage{pgf-pie}
\usepackage{enumitem}

\newcommand\todoEB[1]{\textcolor{red}{EB: #1}}
\DeclareUnicodeCharacter{202C}{}

\def\BibTeX{{\rm B\kern-.05em{\sc i\kern-.025em b}\kern-.08em
    T\kern-.1667em\lower.7ex\hbox{E}\kern-.125emX}}
\begin{document}

\title{
\footnotesize
\framebox[1.01\width]{\parbox{\dimexpr\linewidth-2\fboxsep-2\fboxrule}{If you cite this paper, please use the PEMWN 2020 reference: G. Restuccia, H. Tschofenig, E. Baccelli. Low-Power IoT Communication Security: On the Performance of DTLS and TLS 1.3. In Proc. of 9th IFIP/IEEE PEMWN, Dec. 2020.}}
 \ \\ \ \\ \ \\
\Huge Low-Power IoT Communication Security:\ \\
On the Performance of DTLS and TLS 1.3\ \\
\thanks{This work was partly funded by H2020 SPARTA.}
}



\author{\IEEEauthorblockN{Gabriele Restuccia}
\IEEEauthorblockA{\textit{CNIT \& Freie Univerist\"at Berlin}}
\and
\IEEEauthorblockN{Hannes Tschofenig}
\IEEEauthorblockA{\textit{Arm Ltd.}}
\and
\IEEEauthorblockN{Emmanuel Baccelli}
\IEEEauthorblockA{\textit{Inria \& Freie Univerist\"at Berlin}}
}

\maketitle

\begin{abstract}
Similarly to elsewhere on the Internet, practical security in the Internet of Things (IoT)
is achieved by combining an array of mechanisms, at work at all layers of the protocol stack, in system
software, and in hardware. Standard protocols such as Datagram Transport Layer Security (DTLS 1.2) and Transport Layer Security (TLS 1.2) are often recommended to secure communications to/from IoT devices. Recently, the TLS 1.3 standard was released and DTLS 1.3 is in the final stages of standardization. In this paper, we give an overview of version 1.3 of these protocols, and we provide the first experimental comparative performance analysis of different implementations and various configurations of these protocols, on real IoT devices based on low-power microcontrollers. We show how different implementations lead to different compromises. We measure and compare bytes-over-the-air, memory footprint, and energy consumption. We show that, when DTLS/TLS 1.3 requires more resources than DTLS/TLS 1.2, this additional overhead is quite reasonable. We also observe that, in some configurations, DTLS/TLS 1.3 actually decreases overhead and resource consumption. All in all, our study indicates that there is still room to optimize the existing implementations of these protocols.

\end{abstract}

\section{Introduction}\label{sec:introduction}

As the Internet of Things (IoT) deployment unfolds, the list of attacks against IoT devices gets longer.
To address the security concerns,
governmental institutions, industry groups, and researchers have been publishing recommendations and guidelines for securing IoT devices. 
Examples include the `Baseline Security Recommendations for IoT'~\cite{ENISA} published by the European Union Agency for Cybersecurity (ENISA) and the `Core Cybersecurity Feature Baseline for Securable IoT Devices: A Starting Point for IoT Device Manufacturers'~\cite{NIST} published by 
the National Institute of Standards and Technology (NIST).
Renowned security expert Bruce Schneier~\cite{Schneier} published a long list of available guidelines and it is immediately clear that there is no shortage on advice for IoT manufacturers and embedded developers.

While these sets of guidelines vary in the level of detail, they all ask for communication security to be integrated into IoT devices. 
Over the years, Transport Layer Security (TLS) has become the most widely used (and the most carefully scrutinized) approach to secure data in transit over the Internet.
To no surprise, some IoT security guidelines actually mandate the use of TLS to provide end-to-end communication integrity, confidentiality and authenticity.
Version 1.2 of TLS~\cite{RFC7925} has so far been the recommended standard for securing protocols operating over TCP, along with its UDP-based variant Datagram Transport Layer Security (DTLS).

Recently, TLS 1.3~\cite{RFC8446} was published, and DTLS 1.3 is in the final stages of standardization.
A reasonable question is thus: How appropriate is TLS 1.3 for IoT?
In particular, we focus here on resource-constrained IoT devices, such as described in RFC 7228~\cite{RFC7228}. These devices are based on microcontrollers -- for instance Arm Cortex M-based  microcontrollers -- on which run real-time operating systems, such as FreeRTOS, RIOT, Micrium's $\mu$C/OS, Mbed OS, among others~\cite{hahm2016surveyOS}. 
Compared to computers that run full-blown operating systems, such as Linux, constrained IoT devices use a fraction of the power and are equipped with RAM and Flash sizes in the kilobyte range.

{\bf Related work --} 
There is so far surprisingly little work evaluating DTLS/TLS 1.3 on IoT hardware.
The only prior work we know of is \cite{bossi2016tls13}, which evaluates a preliminary draft version of TLS 1.3 on STM32-based microcontrollers.
Compared to \cite{bossi2016tls13}, our paper analyzes the final TLS 1.3 standard, as well as the latest version of DTLS 1.3.

Open source implementation of DTLS 1.3 we know of so far are limited to a partial implementation in Go ~\cite{Mint-DTLS} and a prototype in Mozilla NSS~\cite{NSS}, neither of which fit in an embedded network stack applicable to microcontrollers.
To the best of our knowledge, this paper presents the first experimental study evaluating and comparing DTLS/TLS 1.2 with DTLS/TLS 1.3, on low-power microcontrollers.

{\bf Paper contributions --}
The main contributions of this paper are the following:
\begin{itemize}
\item we provide a comparative overview of DTLS/TLS 1.3 and 1.2 protocol specifications;
\item we provide the first open source implementation of the DTLS 1.3 protocol targeting small embedded systems;
\item we conduct experiments on IoT microcontrollers, comparing the performance of different configuration and multiple implementations of DTLS/TLS 1.3 and 1.2;
\item we measure and compare the footprint of secure communication solutions in terms of bytes over-the-air, RAM and Flash memory requirements, energy consumption;
\item we overview next steps and optimization potential towards lower footprint with IoT communication security at the transport layer and above.
\end{itemize}

{\bf Paper organization --} 
This paper is organized as follows:
In Section~\ref{sec:background}-\ref{sec:overview} provides necessary background on TLS, DTLS and overviews protocol differences between 1.3 and 1.2 versions.
In Section~\ref{sec:test-setup} we describe our experimental test setup. 
In Section~\ref{sec:evaluation} we present the results of our comparative performance evaluation, based on memory, traffic overhead, and energy measurements. 
Then, in Section~\ref{sec:future} we discuss perspectives, before we conclude. 
\ \\

\section{Background}\label{sec:background}

\begin{figure}[!t]
\centering
\includegraphics[width=8.5cm]{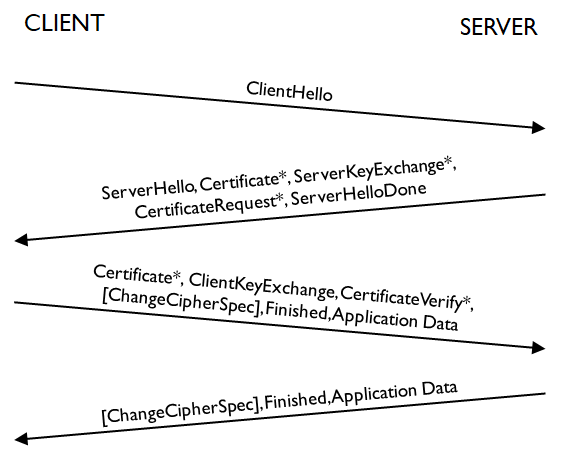}
\caption{Full TLS 1.2 Handshake. Notation: * optional field, [] piggybacked message not part of handshake per se.}
\label{fig:tls12-hs}
\end{figure}

TLS is a communication security protocol standardized by the Internet Engineering Task Force (IETF).
Provided two end points A and B can communicate over a network,
TLS establishes and maintains a secure communication channel between A and B.
TLS guarantees the integrity of the data flowing through this channel, as well as its authenticity, and its confidentiality.
\begin{description}[leftmargin=0pt]
\item[TLS Handshake.]
In a first phase, called the \emph{handshake layer}, TLS specifies an authentication and key exchange protocol.
The handshake results in the authentication of A and B (using asymmetric or pre-shared keys) and the establishement of a master secret known only to A and B: a symmetric key.
In this context, the end point initiating the handshake is called the client, while the other end point is called the server.
Fig. \ref{fig:tls12-hs} shows client-server control message exchanges for a TLS (1.2) handshake.
During the handshake, prior to establishing the master secret, client and server automatically negotiate the cipher suite parameters to be employed -- as several configurations are specified by the protocol.
\item[TLS Record Layer.]
In a second phase, called the \emph{record layer}, the parameters and the symmetric key established by the handshake layer are used to encrypt and decrypt data (i.e. so-called records) sent from A to B, and vice-versa, ensuring confidentiality, integrity and authenticity.
This key is used until the channel gets torn down or a maximum limit of records have been exchanged. The maximum limit depends on the algorithm. Beyond that point, should A and B need to communicate securely again, a new handshake is performed.
\item[TLS versus DTLS and ATLS.]
TLS was intially designed to operate at the transport layer, on top of a transport protocol that guarantees reliable and in-order delivery of data (typically a TCP connection).
TLS provided a widely adopted communication security solution for connection-oriented transport protocols but there was also a need to offer communciation security for connectionless transports, such as UDP. The resulting design was published as Datagram TLS (DTLS~\cite{RFC7925}). The DTLS design intentionally makes as few modifications to TLS as possible, 
which enables large code re-use. It was, however, necessary to include enhancements to the handshake protocol to detect message loss and duplication. It was also necessary to add a DoS protection feature with an optional cookie exchange. 
While TLS and DTLS are typically used on top of the transport layer the protocol is flexible enough to be applied to run over other layers in the protocol stack as well. For instance, Application TLS (ATLS~\cite{I-D.friel-tls-atls}) uses TLS and DTLS over an application layer protocol, such as CoAP or HTTP. In network access authentication TLS has been encapsulated in the Extensible Authentication Protocol for use over link layer protocols, such as IEEE 802.1X, and the AAA infrastructure~\cite{RFC5247}.
\item[Other IoT Communication Security Approaches.]
Aside of TLS and DTLS, alternative communication security solutions have also emerged.
The most prominent alternatives targeting IoT are based on COSE~\cite{RFC8152} and OSCORE~\cite{RFC8613},
which aim at securing transferred data higher up in the protocol stack, i.e. at the application layer and at the CoAP layer, respectively.

Object Signing and Encryption (COSE) is a signing and encryption scheme which is conceptually similar to JavaScript Object Signing and Encryption (JOSE).
COSE can be used in "standalone" fashion to protect data transmitted over the network. 
For instance, a COSE usecase is to protect a firmware update binary (or more precisely, for signing metadata about the firmware and, optionally, encrypting the firmware image~\cite{DBLP:journals/access/ZandbergSPTB19}). The COSE specification contains a range of security primitives and application developers would most likely select the mechanisms they are most interested in rather than using the complete specification.

Object Security for Constrained RESTful Environments (OSCORE) reuses COSE to protect some (but not all) CoAP headers and the CoAP payload. 

Since OSCORE does not offer key management itself, it has to rely on a separate key management protocol. 
Key management protocols currently investigated in the IETF include Ephemeral Diffie-Hellman over COSE (EDHOC) and ATLS. 
EDHOC rebuilds a handshake protocol based on COSE while ATLS uses TLS and DTLS at the application layer. 
\item[Overhead \& Performance of DTLS and TLS.]
Optimizing the footprint of (D)TLS is important on low-end IoT devices. For instance, our measurements show that, for a simple IoT firmware using a 6LoWPAN/CoAP network stack, transport layer security code adds up to 50\% of the total binary size (Flash memory), depending on the implementation and the TLS configuration. In terms of RAM, some (D)TLS configurations consume 20\% of the total RAM usage of the firmware.

TLS relies on heavy computations happening in the handshake, based on assymmetric crypto.
Hence, the less frequent handshakes are, the better (subject to security still being guaranteed).
Comparatively, the record layer is triggered more often, but incurs computations based on symmetric crypto which are more lightweight.

As shown in Fig. \ref{fig:tls12-hs}, to complete the handshake, control data exchanged by client and server requires several round-trip times (RTT), thus incurring some delay and some communication overhead to setup the secure channel.
Obviously, the quicker the handshake is over, with the less bits transmitted, and the less memory required, the better (subject to security still being guaranteed).

DTLS/TLS 1.3 brings a number of qualitative improvements over DTLS/TLS 1.2, which we overview next. 
A subsequent question is: at which price come these improvements?
We will thus next compare DTLS/TLS 1.3 with DTLS/TLS 1.2 from a quantitative, performance point of view.
\end{description}

\section{From TLS/DTLS 1.2 to TLS/DTLS 1.3: an Overview}\label{sec:overview}

First we overview the breadth changes between TLS/DTLS 1.2 and TLS/DTLS 1.3.
Then, we illustrate the handshake protocols for TLS and DTLS, we illustrate the use of the Connection ID, and explain improvements to the record layer and discuss backward compatibility.

\subsection{From TLS 1.2 to TLS 1.3}
\paragraph*{Shorter Handshakes} 
The number of roundtrips have been reduced. The newly introduced 0-RTT exchange allows application data to be transmitted already with the first message. As a result, replay protection cannot be offered by the TLS layer for this so-called early data and it has to be provided by the application layer. Using a regular exchange a client can send application data after the first roundtrip. To convey the necessary parameters for setting up the security context the client has to add extra information to the ClientHello. The reduction of latency is made possibly by making additional assumptions about what knowledge the client has about the server. It is assumed that the client knows the server, or at least keeps some information about the server cached, and configuration information (such as algorithms) rarely changes. 

\paragraph*{More Hurdles for Traffic Analysis}
Privacy protection has been improved by encrypting handshake messages as early as possible. To make traffic analysis more difficult, record layer padding was added. 

Increased confidentiality protection is offered with the mandatory use of perfect forward secrecy for public key-based key exchange modes. As a consequence, the RSA key transport mechanism was removed and this decision was met with considerable skepticism by enterprise network operators and firewall vendors since it limits their deep packet inspection abilities. 

\paragraph*{Ciphersuites Cleanup}

The ciphersuite concept was re-visited: algorithm negotiation is untangled from the authentication and key exchange mechanism. The ciphersuite list was cleaned-up as well. 

The three modes based on pre-shared keys (PSK) present in TLS 1.2, namely PSK-based ciphersuites, classical TLS session resumption, and session resumption without server-side state, have been merged into a single mode. 

Support for Password-Authenticated Key Agreements (PAKE)s, which were available in TLS/DTLS 1.2 with SRP~\cite{RFC5054} and J-PAKE~\cite{I-D.cragie-tls-ecjpake}, were removed in TLS 1.3~\footnote{The Crypto Forum Research Group (CFRG) is currently conducting a PAKE selection process and the integration of the PAKE algorithm into TLS 1.3 is one of the desirable features. Hence, it can be expected that sooner or later a standardized PAKE algorithm will be available for use with TLS/DTLS 1.3.}.

Authenticated encryption with additional data (AEAD) has become the go-to-choice for symmetric encryption algorithms. This decision reflects a trend in the TLS working group (and in the industry in general) where other modes of operations for ciphers have either been deprecated (such as RC4) or discouraged (e.g. AES with CBC).

\paragraph*{Revised Key Hierarchy}
The key derivation algorithm now makes use of the HMAC-based Extract-and-Expand Key Derivation Function (HKDF) construct defined in RFC 5869~\cite{RFC5869} in a newly specified key hierarchy,  which enables a systematic analysis of the handshake using formal methods, leveraging the independence of the different keys (see Section 7 of \cite{RFC8446}). From an implementation point of view, this change led to a larger code size compared to earlier TLS versions, which used a relatively simple key hierarchy. 

\paragraph*{Changes Concerning Negotiation}
Renegotiation was re-defined with the introduction of a post-handshake authentication mechanism. 
Furthermore, version negotiation was rationalized. 

\paragraph*{Removed Functionalities}
Custom Diffie-Hellman groups were removed due to unnecessary complexity. 
Compression functionality was removed because of attacks like CRIME, TIME and BREACH~\cite{RFC7457}.

\subsection{From DTLS 1.2 to DTLS 1.3}

DTLS 1.3 re-uses, like earlier versions, much of the TLS 1.3 handshake. There are, however, new features integrated that did not exist in previous versions, such as the Connection Identifier and the record layer format was heavily optimized. The mechanism for ensuring the reliability of handshake messages was also revisited to improve performance of the handshake in networks with a high packet loss rate. 

\paragraph*{Connection Identifier (CID)}
In DTLS, the source IP address and source port of the sender is used to identify a security context. With CIDs a new identifier is carried in the record layer of encrypted DTLS packets, which is used for the lookup of established security state. With this extension an established security context can be used significantly longer without the need to re-run the full handshake or a resumption handshake. This can lead to significant performance improvements. CIDs are particularly useful in an IoT context where devices sleep for extended time periods and likely run into the problem of expired state at Network Address Translators (NATs). Note that this extension has also been backported to DTLS 1.2~\cite{I-D.ietf-tls-dtls-connection-id} but the DTLS 1.2 CID solution it offers different privacy characteristics.

\paragraph*{Record Layer Optimization}
Once the handshake completes, application data is protected using the established keys. Every payload is wrapped inside a record layer structure. To reduce the overhead of the record layer structure, the DTLS 1.3 specification introduced a variable length format and removed or shortened several fields. A sequence number encryption scheme has been added to improve privacy protection by limiting the ability correlation by an on-path adversary.

\paragraph*{Redefined Retransmission Mechanism}
Prior to DTLS 1.3 the retransmission granularity was at the level of an entire flight, i.e. a series of message. In DTLS 1.3 an explicit ACK message has been added allowing the explicit acknowledgement of individual handshake messages.

\begin{figure}[!t]
\centering
\includegraphics[width=8.5cm]{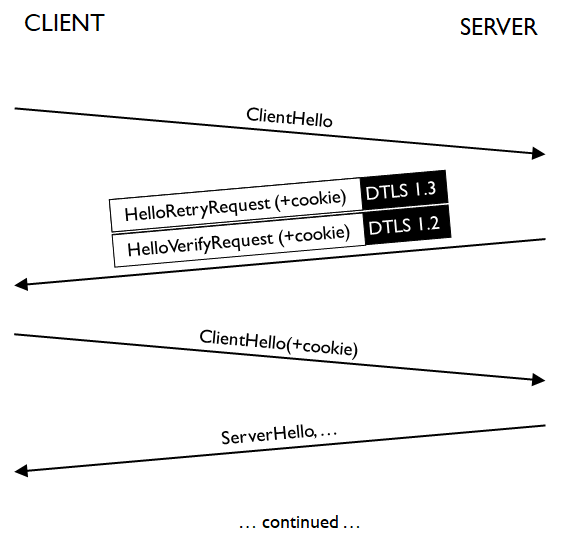}
\caption{DTLS 1.2 and DTLS 1.3: DoS Mitigation Technique.}
\label{fig:dtls-hs}
\end{figure}

\paragraph*{DoS Protection with TLS 1.3 HelloRetryRequest}
DTLS 1.2~\cite{RFC6347} required an enhancement of the handshake for denial of service protection to avoid reflection attacks and resource exhaustion on the server side. This is accomplished via the introduction of a dedicated message in earlier versions of DTLS, namely the HelloVerifyRequest. In DTLS 1.3 this message has been replaced by the HelloRetryRequest since this message has already been introduced in TLS 1.3, as shown in Fig. \ref{fig:dtls-hs}.

\subsection{TLS 1.3 Handshake Analysis}
In the rest of this section we use the following notation borrowed from the TLS 1.3 specification: `+  indicates a noteworthy extension', `*' refers to an optional message orextension, `[]', `()', and `\{\}' show encrypted messages whereby the keys used to encrypt these messages differ. 

When comparing the TLS 1.2 with the TLS 1.3 handshake it is obvious that a few messages have been removed, namely the ServerHelloDone, the ChangeCipherSpec, the ServerKeyExchange, and the ClientKeyExchange. In the TLS 1.2 handshake the Finished message is the first (and only) encrypted handshake message. The contents of the ServerKeyExchange and ClientKeyExchange messages varies with the negotiated authentication and key exchange method. With TLS 1.3 its content has been moved to extensions in the ClientHello and ServerHello messages. The ServerHelloDone and the ChangeCipherSpec messages were removed without replacement. 

\begin{figure}[!t]
\centering
\includegraphics[width=8.5cm]{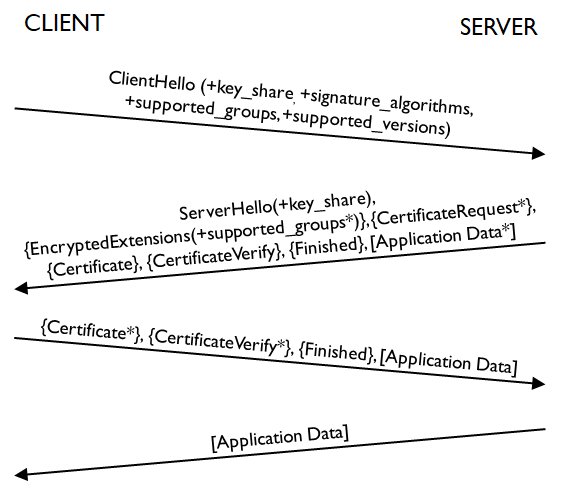}
\caption{TLS 1.3 Public Key based Authentication.}
\label{fig:tls-pk-hs}
\end{figure}

Looking at TLS 1.3 the public key-based authentication mode is probably the most important and it is shown in Figure~\ref{fig:tls-pk-hs}. It always uses perfect forward secrecy using an ephemeral Diffie-Hellman exchange. Figure~\ref{fig:tls-pk-hs} shows the extensions in the ClientHello that have to be used with this mode. Application data is already sent by the client after the first roundtrip and the EncryptedExtensions provides confidentiality protection of many of the extensions carried in the ServerHello in earlier versions of TLS. If mutual authentication is desired, which is common in IoT deployments, then the server uses the CertificateRequest message. The Certificate, CertificateVerify and the Finished message have kept their semantic of earlier TLS versions. Size-wise the biggest payloads in this exchange are the Diffie-Hellman public keys exchanged between the two parties and the certificates. 

\begin{figure}[!t]
\centering
\includegraphics[width=8.5cm]{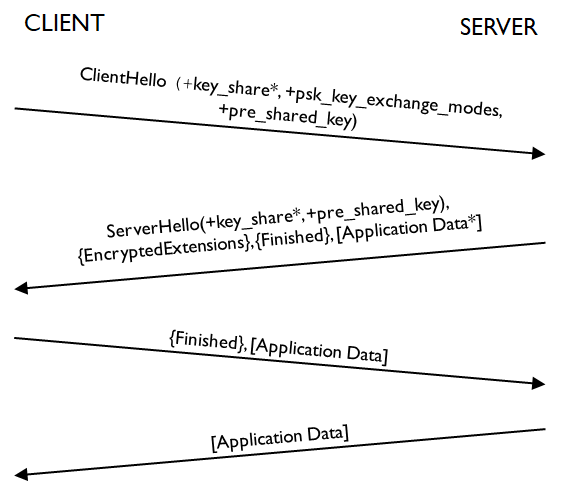}
\caption{TLS 1.3 Pre-Shared Key (PSK) Authentication.}
\label{fig:tls-psk-hs}
\end{figure}

The pre-shared key (PSK) authentication mode is shown in Figure~\ref{fig:tls-psk-hs} and it optionally uses a Diffie-Hellman exchange (with the Diffie-Hellman public keys conveyed in the key\_share extension). This mode harmonizes three PSK variants available in previous TLS versions into one. The pre\_shared\_key extension in the ClientHello signals the use of the PSK mode and contains the PSK identity and the so-called binder. The PSK identity identifies the shared secret and the binder is a structure that contains one or multiple MAC values. Application data is already sent by the client after the first roundtrip. In this mode many of the messages shown in the public key-based mode are missing. In general, this mode is computationally much less demanding, leads to shorter message sizes, and requires less code size because in the simplest configuration it does not require any asymmetric crypto operations. 

\begin{figure}[!t]
\centering
\includegraphics[width=8.5cm]{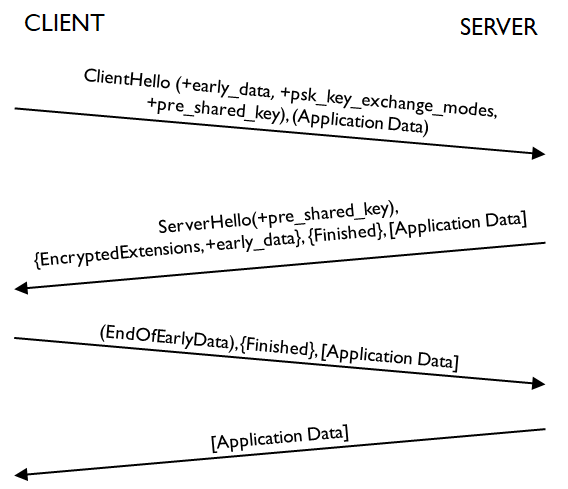}
\caption{TLS 1.3 0-RTT Data.}
\label{fig:tls-0rtt}
\end{figure}

The 0-RTT mode, shown in Figure~\ref{fig:tls-0rtt}, allows a client to send application data, called early data, already with the first message. This mode represents a new feature in TLS 1.3 and does not have an equivalent in earlier TLS versions. It does, however, introduce the risk of replay attacks and an application is therefore required to check for replay attacks. The use of early data likely requires API extensions in the embedded TLS stack. The 0-RTT mode makes use of a PSK established in an earlier exchange (or an externally configured PSK) and can therefore be seen as an optimized version of the PSK authentication mode. 

Table~\ref{tab:performance} shows a ranking of the authentication modes indicating when the client send application data and what the expected over-the-wire message size is. Note that there is a tradeoff between over-the-wire performance, functionality and code size/RAM utilization: smaller packet sizes are typically accomplished with extra performance optimizations, which either require more complex code or additional RAM (or in some cases both). Note that some of the extensions developed for earlier TLS versions can still be used with TLS 1.3, such as raw public keys (RPKs)~\cite{RFC7250} and TLS cached info~\cite{RFC7924}. These extensions have not been taken into account during the performance measurement and hence there is room for further optimizations. 

\begin{table}[]
\begin{tabular}{|l|l|l|}
\hline
Mode &  App data can be sent & Data Size \\ \hline
PK mode (+ X.509 certs) & after 1st RT & ++++ \\ \hline
PK mode (+ RPKs)  & after 1st RT & +++ \\ \hline
PSK (+PFS)  &  after 1st RT & ++ \\ \hline
Plain PSK & after 1st RT & + \\ \hline
0-RTT & with 1st msg & + \\ \hline
\end{tabular}
\caption{Performance of the different TLS 1.3 authentication modes.}
\label{tab:performance}
\end{table}

\subsection{DTLS 1.3 Backward Compatibility}
Introducing protocol changes to improve efficiency is relatively easy from an engineering point of view until the deployment reality settles in. Engineers have been exposed to protocol ossification, which is the result of software and hardware deployed in the middle of the network.  Unfortunately, it also impacted the design of TLS 1.3 in a subtle way. With the changes to various handshake messages TLS 1.3 deployment experiments have shown problems with failed connection attempts and even with blacklisted IP addresses resulting from misbehaving firewalls and intrusion detection systems. As a consequence, the TLS working group descided to modify TLS 1.3 at a large stage in the standardization process such that it looks on the wire like TLS 1.2 session resumption. This required the newly introduced HelloRetryRequest message to look like the ServerHello message (which is bigger on the wire), and to exchange fake ChangeCipherSpec messages. Luckily, DTLS 1.3 does not suffer from the same level of ossification and changes to the handshake messages and even to the record layer are still possible to deploy. 

\subsection{DTLS 1.3 Record Layer}

\begin{figure}[!t]
\centering
\includegraphics[width=8.5cm]{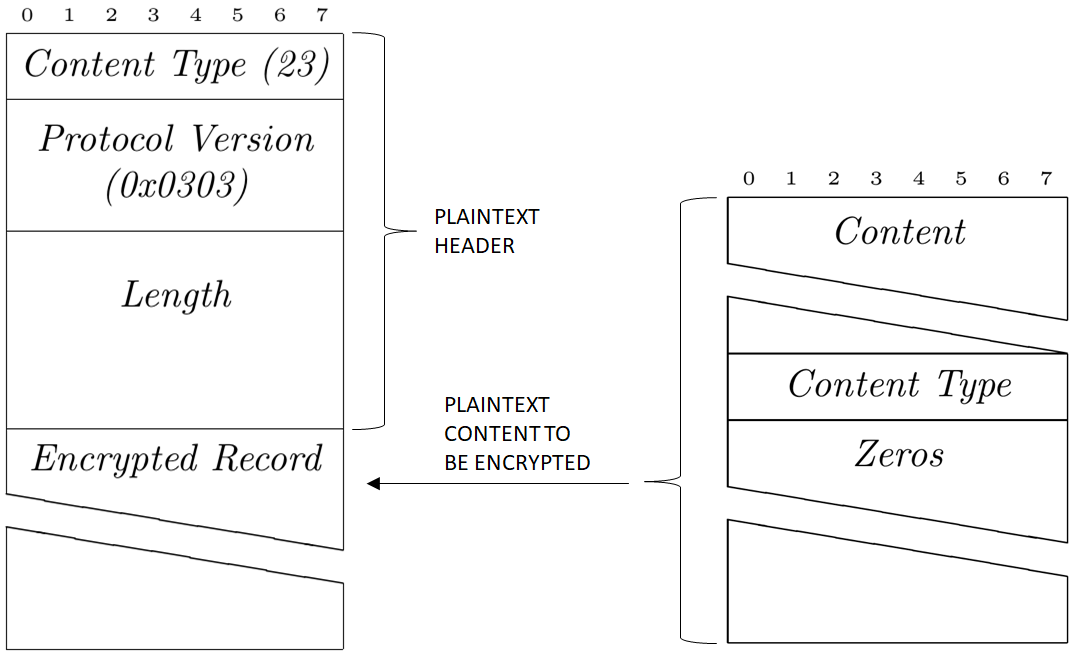}
\caption{TLS 1.3 Record Layer Structure.}
\label{fig:TLS13_RecordLayer_Complete}
\end{figure}

The TLS 1.3 Record Layer format has changed to offer better privacy protection (and due to backwards compatibility reasons) by faking the outermost content type and protocol version fields. Figure~\ref{fig:TLS13_RecordLayer_Complete} shows the nested structure with the true content type value encapsulated inside the encrypted payload. Note also the additional zero byte padding following the plaintext content.  

\begin{figure}[!t]
\centering
\includegraphics[width=3cm]{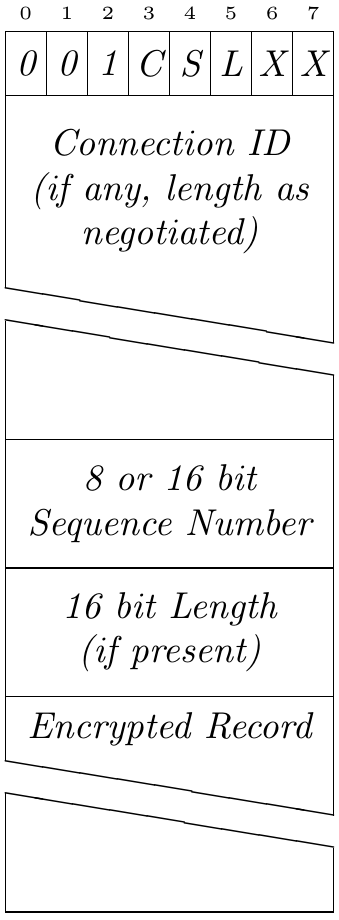}
\caption{DTLS 1.3 Ciphertext Structure.}
\label{fig:DTLS13_Ciphertext_Structure_Full}
\end{figure}

In contrast, the DTLS 1.3 record layer structure has been highly optimized with a variable length encoding. The new structure is shown in Figure~\ref{fig:DTLS13_Ciphertext_Structure_Full} the Connection ID, and Length field being optional. The presence or absence of the fields as well as the length of the sequence number is controlled by bitmask. In the most minimalistic case the DTLS 1.3 record layer only contains the 8 bit long bitmask, 8 bits of sequence numbers followed by the encrypted record whereby in this case the length information has to be retrieved from the underlying transport layer. 

\subsection{Summary of Protocol Comparison}
Comparing TLS/DTLS 1.3 to 1.2, the list of new features and changes is quite long. 
The primary improvements we observe, protocol-wise, in 1.3 compared to 1.2, are in terms of:
\begin{itemize}
\item added security guarantees, and
\item minimized setup delay until the point where application data can first be sent.
\end{itemize}

Nevertheless, taking account all of the changes (also including feature removals),
it is not straightforward to gauge the impact of such changes on performance metrics such as 
code size on micro-controllers, 
the overall required amount of cryptographic operations,
or the total control data overhead.

Towards a more precise assessment, the next sections thus present experiments we carried out on IoT hardware.
\section{Experimental Setup}
\label{sec:test-setup}
In this section we provide an overview of our experimental setup. 
\ \\
\paragraph*{IoT Hardware Setup}
Our experiments were carried out on three Arm Cortex M-based IoT devices: 
\begin{itemize}
\item the {\bf nRF52840-DK} uses a commercially available Arm Cortex-M4 microcontroller from Nordic Semiconductor, which has a generous 1MB of Flash memory and 256kB of RAM. This was our primary hardware. 
\item the {\bf IoT-Lab M3 Node} uses a custom-designed development board that includes a Arm Cortex-M3 microcontroller from STMicroelectronics, which has 512kB of Flash memory and 64kB of RAM, and utilizes the STM32F103REY MCU. This board has been developed for the open access testbed FIT IoT-Lab~\cite{2015-IoT-Lab}. This board was used to determine the impact of the use of a Cortex M3 on the flash footprint. 
\item the {\bf Keil MCBSTM32F400 evaluation board} is based on an Arm Cortex-M4 from STMicroelectronics (STM32F407IG), which has 1MB of Flash memory and 192 kB of RAM. This board was used to verify the flash, heap, and energy measurements. 
\end{itemize}

For reliable measurements of the communication we used 6LoWPAN over Ethernet over serial. 

\ \\
\paragraph*{Software and Network Setup}
We provide the first implementation of DTLS 1.3 targeting low-power microcontrollers, open source published at~\cite{Hannes-DTLS13}.
For our setup, we used RIOT~\cite{2018-RIOT} as the real-time operating system in version 2019.10, which includes networking middleware.
Our setup uses two RIOT nodes communicating via a 6LoWPAN/CoAP protocol stack, including support for different DTLS/TLS libraries and configurations.

Leveraging these software foundations, we then evaluate different configurations, namely: 
\begin{itemize}
\item {\bf TLS/DTLS protocol versions}: TLS 1.2 and 1.3, DTLS 1.2 and 1.3;
\item {\bf authentication and key exchange profiles}: PSK-based authentication and AES128/AES256 on one hand, and profiles with ECDSA-ECDHE-based authentication (using the NIST curves) and AES128/AES256 on the other hand;
\item {\bf key size variations}: symmetric cryptography with 128-bit and 256-bit for AES and the corresponding key size for \\ECDHE/ECDSA with NIST P256r1 and P521r1 curves. 
\item {\bf embedded TLS/DTLS stack implementations}: we use open source implementations from WolfSSL (version 4.4.0), and Mbed TLS (version 2.16), as well as our own TLS/DTLS 1.3 implementation.
\end{itemize}
We wrote an application to test the different configurations and the code is published at~\cite{Gabriele-RIOT-branch} for reproduceability. Our application uses TLS and DTLS on top of CoAP, as specified in~\cite{I-D.friel-tls-atls}. Securing the communication channel end-to-end above CoAP is indeed a trend in IoT, as also demonstrated by alternative standards such as OSCORE~\cite{RFC8613}.
We thus chose this configuration as it gives a good base to pit variants of TLS and DTLS both against one another, and with such alternatives (in future work).
\ \\

To perform our energy measurements, we primarily used the {\bf Nordic nRF6707} power profiling kit, which was designed to work with the nRF52840-DK. The measurement captures the entire protocol exchange, including RTOS handling, network communication, processing by the networking stack, as well as the entire TLS/DTLS handshake. We start the measurement when the first packet is sent out to initiate the TLS/DTLS handshake and stop it when we close the TLS/DTLS handshake. Because we use ATLS, the CoAP transmission as our transport for TLS/DTLS is included in the energy measurements. For verification of the measurements we also used the Keil ULINK plus together with the MCBSTM32F400 evaluation board.

On the IoT-Lab M3 Node and on the nRF52840-DK we utilized the GNU Arm Embedded Toolchain (9-2020-q2-update) with GCC 9.3.1. The compiler has been configured to optimize for code size. For the Keil MCBSTM32F400 we used the Keil uVision 5 IDE in version 5.26.2.0. As a compiler we used Arm Compiler version 5 with the highest optimization level (level 3). The linker was configured with data compression
disabled. We used a scatter file to influence the placement of the different components in RAM and flash memory.

\paragraph*{Metrics}
To comparatively evaluate the protocols and their implementations, we measure the required 
(i) stack and heap size,
(ii) code size,
(iii) over-the-wire network traffic overhead,
and (iv) energy consumption.

While the absolute numbers are important they ignore the flexibility of embedded TLS stacks whereby the crypto algorithms can be optimized, or replaced by another software implementation. In some cases, depending on the capability of the hardware, the TLS stack can also make use of crypto accelerators in hardware. Note that we did not utilize hardware crypto acceleration provided by the nRF52840-DK. Prior work, such as~\cite{DBLP:journals/access/ZandbergSPTB19}, investigates different embedded crypto libraries and noted the tradeoffs made by different implementations. Hence, we have included tests with different TLS/DTLS implementations.

\section{Experimental Results}
\label{sec:evaluation}
Based on the measurements we present below, we aim to provide some answers to questions that either a specialist, or a neophyte would ask.

\subsection{Impact of protocol versions}
What is the impact of upgrading from 1.2 to 1.3 on resource requirements? 

\paragraph*{Memory footprint.} 
Table \ref{tab:flash:comparison} shows the Flash memory footprint using the Mbed TLS stack. Compared to versions 1.2, these measurements indicate ~20\% increase with TLS / DTLS 1.3. Further measurements on the RAM shown below in Table \ref{tab:RAM:WolfSSL-versus-mbedTLS} indicate no significant impact on stack requirements, and limited impact on peak heap footprint (max 25\% increase for some configurations in one implementation, and for the other implentation, even ~30\% decrease, for ECDHE configurations). 
Our conclusion is that TLS/DTLS 1.3 does not require a substantial amount of additional flash and heap compared to TLS/DTLS 1.2. 
Hence, developers can benefit from (D)TLS 1.3 protocol improvements without the need to upgrade hardware.

\begin{table}
\resizebox{\columnwidth}{!}{
  \begin{tabular}{l l l l}
    \toprule
                                              & 1.2  & 1.3 & Diff \\
    \midrule
     TLS with PSK, AES-128-CCM                & 19004 & 23988 & ‬4,984 \\
     TLS with PSK, AES-256-GCM                & 22942 & 28696 & 5,754 \\
     TLS with ECDHE-ECDSA, AES-128-CCM        & 46892 & 53023 & ‬6,131 \\
     TLS with ECDHE-ECDSA, AES-256-GCM        & 54312 & 61149 & 6,837 \\
     DTLS with PSK, AES-128-CCM               & 22338 & 28136 & 5,798 \\
     DTLS with PSK, AES-256-GCM               & 27320 & 33894 & 6,574 \\
     DTLS with ECDHE-ECDSA, AES-128-CCM       & 50532 & 58097 & ‬7,565 \\
     DTLS with ECDHE-ECDSA, AES-256-GCM       & 57972 & 66215 & 8,243 \\
\bottomrule
\end{tabular}
}
\caption{DTLS/TLS 1.2 vs. DTLS/TLS 1.3 Flash Size (with Mbed TLS on Arm Cortex-M4).}
\label{tab:flash:comparison}
\end{table}

\paragraph*{Bytes over the air.}
Table \ref{tab:wire:comparison} shows some measurements of bytes-over-the air, for various configurations of (D)TLS 1.2 and 1.3. We observe that:
\begin{itemize}
\item for TLS, slighlty increased (~10\% more) bytes-over-the-air, except for PSK AES-256-GCM (+23\%);
\item for DTLS with PSK decreased bytes-over-the-air (~25\% less);
\item for DTLS with ECDHE-ECDSA, decreased bytes-over-the-air (~15\% less).
\end{itemize}

Based on these measurements, we conclude that, for DTLS 1.3 we see a slight decrease of over-the-air overhead, most likely due due to the protocol's record layer optimizations, and a slight increase for TLS 1.3 (overall: max 25\% difference).

\begin{table}
\resizebox{\columnwidth}{!}{
  \begin{tabular}{l l l l}
    \toprule
                                              & 1.2  & 1.3 & Diff \\
    \midrule
     TLS with PSK, AES-128-CCM                & 337 & 380 & 43\\
     TLS with ECDHE-ECDSA, AES-128-CCM        & 1308 & 1371 & ‬63\\
     TLS with ECDHE-ECDSA, AES-256-CCM        & 1454 & 1415 & ‬-39\\
     DTLS with PSK, AES-128-CCM               & 627 & 467 & -160\\
     DTLS with ECDHE-ECDSA, AES-128-CCM       & 1726 & 1500 & ‬-226\\
     DTLS with ECDHE-ECDSA, AES-256-CCM       & 1879 & 1542 & ‬-337‬\\
    \bottomrule
  \end{tabular}
  }
  \caption{Bytes-Over-The Air, measured using Mbed TLS.}
  \label{tab:wire:comparison}
\end{table}

\paragraph*{Energy consumption.}

TLS 1.3 can lead to an energy reduction, as shown in Table \ref{tab:energy:comparison}. The most interesting aspect we observe here is the huge difference between symmetric and asymmetric cryptography: while there is some communication overhead due to the 6LoWPAN protocol stack, the PSK-based exchange consumes a fraction of the energy of a corresponding TLS/DTLS handshake with ECDHE-ECDSA. Hence, it is advisable to avoid using a full ECDHE-ECDSA exchange and instead make use of session resumption or even CID-based exchanges. 

\begin{table}
\resizebox{\columnwidth}{!}{
  \begin{tabular}{l l l l}
    \toprule
                                              & 1.2  & 1.3 & Diff \\
    \midrule
     Mbed TLS - TLS with PSK, AES-128-CCM                &  2.7 &  2.3 & 0.4\\
     Mbed TLS - TLS with ECDHE-ECDSA, AES-128-CCM        & 89.6 & 63.4 & ‬-26.2\\
     Mbed TLS - DTLS with PSK, AES-128-CCM               &  2.0 &  5.3 & 3.3‬\\
     Mbed TLS - DTLS with ECDHE-ECDSA, AES-128-CCM       & 87.5 & 73.3 & ‬-14.2\\
     WolfSSL - TLS with ECDHE-ECDSA, AES-128-CCM    &  76.3    & 77.5  & 1.2\\
     WolfSSL - DTLS with PSK, AES-128-CCM           &   1.9   &  N/A   & N/A\\
     WolfSSL - DTLS with ECDHE-ECDSA, AES-128-CCM   &  77.0   &  N/A   & N/A\\
    \bottomrule
  \end{tabular}
  }
  \caption{Energy Measurements (Millicoulomb).}
  \label{tab:energy:comparison}
\end{table}

Let us nevertheless add a remark on the increase in energy consumption for DTLS 1.2 to 1.3 with PSK-AES-128-CCM, from 2.0 to 5.3 uC. This can be explained by the absence of a feature implemented in the 1.2 version of DTLS, see Mbed TLS version 2.13 \cite{MbedTLS2.13}, that allows multiple DTLS records to be packed into a single datagram (or CoAP message in our case). This reduces the bytes transmitted over the wire considerably because the underlying layers add a lot of header overhead. Needless to say that this feature will also be implemented in the DTLS 1.3 stack in the near future given its benefits. 

\subsection{Impact of implementation}
We were curious as to how different implementations may lead to different tradeoffs. 
To explore this aspect, we examin more precisely different implementations: WolfSSL and mbedTLS.

On one hand, we measured that both implementations consistently require more Flash memory for 1.3 compared to 1.2 (approcimately 20\% more). However, the impact on RAM usage seems to vary depending on the implementation, as shown in Table \ref{tab:RAM:WolfSSL-versus-mbedTLS}.

This data suggests that there is probably room left for optimizations in the implementations. In fact, an optimized version of Mbed TLS exists, as for instance \cite{baremetal} which lowers the RAM requirements down to less than 10 Kb for DTLS with ECDHE-ECDSA with AES-128-CCM by using a different crypto library (TinyCrypt), combined with a more efficient management of send and receive buffers, as well as an improved handling of certificates and of the DTLS retransmission buffers.

\begin{table}
\resizebox{\columnwidth}{!}{
  \begin{tabular}{l l l l l}
    \toprule
          & MbedTLS Heap & MbedTLS Stack & WolfSSL Heap & WolfSSL Stack \\ 
    \midrule
          TLS 1.2 PSK AES-128-CCM & 5749 & 8772 & 3496 & 12 \\
          TLS 1.2 ECC AES-128-CCM & 13879 & 8786 & 7162 & 12 \\
          TLS 1.2 ECC AES-256-GCM & 20603 & 8780 & 7922 & 12  \\
          TLS 1.3 PSK AES-128-CCM & 6757 & 8764 & 6224 & 12 \\
          TLS 1.3 ECC AES-128-CCM & 12914 & 8778 & 9458 & 12 \\
          TLS 1.3 ECC AES-256-GCM & 14366 & 8780 & 10250 & 12  \\
          DTLS 1.2 PSK AES-128-CCM & 5975 & 8772 & 5340 & 12 \\
          DTLS 1.2 ECC AES-128-CCM & 14414 & 8786 & 8540 & 12 \\
          DTLS 1.3 PSK AES-128-CCM & 6934 & 8764 & N/A & N/A \\
          DTLS 1.3 ECC AES-128-CCM & 13248 & 8778 & N/A & N/A \\
    \bottomrule
  \end{tabular}
  }
  \caption{Required RAM in Bytes (Peak Heap \& Stack) for MbedTLS and WolfSSL.}
  \label{tab:RAM:WolfSSL-versus-mbedTLS}
\end{table}

Let us conclude the study of this aspect with two important lessons learnt: 
\begin{itemize}
\item Our experience is that it is quite hard to configue different libraries \texttt{exactly} the same way according to protocol specifications. In particular, we noticed some minor difference in bytes over the wire with the different implementations, which we could not completely eliminate, although we tried our best. Understanding which configuration setting lead to the desired resource utilization tradeoffs tends to be tricky.
\item Depending on the selected crypto library RAM and flash memory requirements vary substantially, as shown in \cite{DBLP:journals/access/ZandbergSPTB19} and in optimization efforts like \cite{baremetal}.
\end{itemize}

\subsection{Impact of underlying hardware}

\paragraph*{Are there differences between microcontrollers?}
 As shown in Table \ref{tab:memory-IoT-lab-M3} there are no significant code size differences on the different the development boards we used. 
 Performance-wise there are, however, differences between microcontrollers as they differ in their speed. 
While crypto code could make use of certain features of microcontrollers, such as the caching capabilities of the Cortex M7 processor, those optimizations are rarely utilized by developers because require a detailed understanding of the processor architecture, expertise in cryptography and time to develop optimized algorithm implementations. It is therefore more likely for developers to make use of the provided hardware crypto acceleration even though we did not update the code to feed it into our tests. Note that there are dedicated benchmarks focused on measuring the performance of MCUs, see \cite{EEMBC}. 

\begin{table}
\resizebox{\columnwidth}{!}{
  \begin{tabular}{l l l l l l l}
    \toprule
                                              & 1.2 Flash  & 1.3 Flash & 1.2 Stack & 1.3 Stack\\ 
    \midrule
     TLS with PSK, AES-128-CCM                & 18938 & 22912 & ‬8772 & 8764  \\
     TLS with ECDHE-ECDSA, AES-128-CCM        & 46752 & 52867 & ‬8786 & 8778  \\
     DTLS with PSK, AES-128-CCM               & 22278 & 28122 & 8772 & 8764  \\
     DTLS with ECDHE-ECDSA, AES-128-CCM       & 50394 & 57973 & ‬8786 & 8778  \\
    \bottomrule
  \end{tabular}
  }
  \caption{Flash \& Stack memory requirements on Arm Cortex-M3 (with Mbed TLS).}
  \label{tab:memory-IoT-lab-M3}
\end{table}

Note: Fig. \ref{tab:RAM:WolfSSL-versus-mbedTLS} indicates that all configurations of Mbed TLS require an 8 kB stack. This is the result of configuration settings for the AES library where developers can decide whether they want to use more RAM and/or flash for the benefit of improved performance. In our configuration we decided not to place the tables in flash nor to use smaller tables. Hence, we are experiencing a hit on the stack size. Setting the configuration option for smaller tables cuts the stack size requirement down to 2600 bytes.

\paragraph*{Availability of hardware crypto support.}
We did not do any measurments to illustrate this point because available hardware crypto is typically not supported out-of-the-box by popular implementations. However, making good use of such hardware capabilities could change the equation. The EEMBC published SecureMark-TLS benchmark~\cite{Securemark} tried to answer the question about the performance of IoT hardware with and without hardware crypto support.

\subsection{Impact of crypto suite}

\paragraph*{Differences between PSK and ECDHE.}

To no surprise, the overhead of ECDHE compared to PSK is big. But how big?  Fig. \ref{tab:PSK-versus-ECDHE-Flash} shows ECDHE needs approx. 30 kB more Flash memory, which means an increase of roughly ~100\% in code size compared to PSK.

In terms of RAM, measurements shown Fig. \ref{tab:RAM:WolfSSL-versus-mbedTLS} indicate for instance that TLS 1.2 with ECDHE-ECDSA AES-256 peaks at ~15kB more (which means roughly 3.5 times more heap required) compared to TLS1.2 with PSK AES-128, with the MbedTLS implementation.

Last but not least, in terms of energy, measurements shown Table \ref{tab:energy:comparison} indicate that DTLS 1.2 with ECDHE-ECDSA AES-128 burns more than 40 times more energy compared to DTLS 1.2 with PSK AES-128 (measured with Mbed TLS).

\begin{table}
\resizebox{\columnwidth}{!}{
  \begin{tabular}{l l l l l l l}
    \toprule
	                                      & PSK AES128  & PSK AES256 & ECDHE AES128 & ECDHE AES256 \\ 
    \midrule
	  TLS 1.2       & 19kB & 24kB & 46kB & 54kB  \\
          TLS 1.3       & 23kB & 28kB & 53kB & 61kB  \\
    \bottomrule
  \end{tabular}
  }
  \caption{Flash memory requirements for mbedTLS on an ARM Cortex-M4.}
  \label{tab:PSK-versus-ECDHE-Flash}
\end{table}

Next Fig. \ref{pie:flash:PSK} breaks down the relative flash memory footprint per component, for TLS 1.3 PSK with AES-128-CCM. Note that the code size needed for symmetric crypto is almost as large as the code size for the TLS protocol itself! 
Hence, the crypto libraries in use (and the cipher suite) have a major impact on the (D)TLS footprint, which should \emph{not} be underestimated.
More in detail, we distinguish four components:
(i) HKDF code, to provide this functionality as specified in RFC 5869 and in the TLS 1.3 specification, 
(ii) the Random Number Generation (RNG),
(iii) Symmetric crypto code, which includes SHA256, AES, CCM, and the wrapper functions for more convenient use of cryptographic primitives, and finally
(iv) TLS code, including the client-side code and code shared by both client- and server-side, as well as code for managing ciphersuites.

\begin{figure}
\resizebox{\columnwidth}{!}{
\begin{tikzpicture}
 \pie [polar, explode=0.1, text=legend]
    {33/Asymmetric Crypto,
     22/TLS,
     21/X.509,
     18/Symmetric Crypto,
     3/HKDF,
     3/RNG}
\end{tikzpicture}
}
\caption{Relative sizes of Flash memory per component, for TLS 1.3 ECDSA-ECDHE (P2561) with AES-128-CCM.}
\label{pie:flash:ECDHE}
\end{figure}
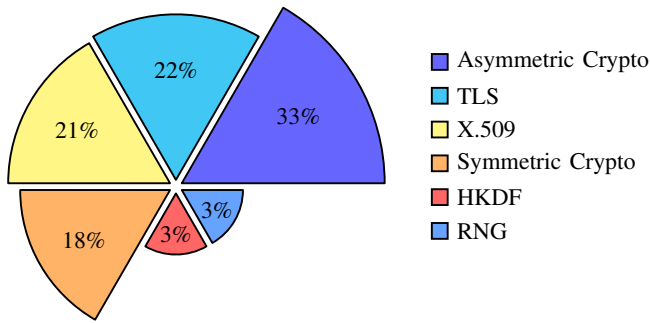

In comparison, Fig. \ref{pie:flash:ECDHE} shows that this footprint changes substantially for ECDHE with AES-128-CCM. 
In this configuration more than 70\% of the memory footprint is taken by crypto (among which 50\% for asymmetric crypto alone, counting also X.509). More in detail, we distinguish two additional components compared to Fig. \ref{pie:flash:PSK}, namely:
(v) asymmetric crypto code, including code for the bignum library and functionalities needed for ECC, ECDSA, ECDHE, and
(vi) X.509 code, including ASN1 libraries, public key processing functions and various support functions.

\paragraph*{Differences between AES128/256, SHA256/384, bigger curves.}
Based on our aformentioned measurements 
we observe that the differences between these cipher suites are
\begin{itemize}
\item negligible concerning Flash footprint;
\item noticeable concerning RAM (heap);
\item major in terms of energy consumption.
\end{itemize}

\paragraph*{Differences between CCM and GCM.}
Based on our measurements, we observe that the main impact of GCM versus CCM is in RAM footprint (i.e. in terms peak heap demand).

\subsection{Impact of the protocol}
A surprising observation is the relatively small difference in terms of bytes-over-the-air with DTLS compared to TLS, in particular for ECC. 
This is especially surprising because DTLS is the traditional solution on low-power wireless, where "every sent byte counts".
A big contributor to the total size of the exchanged data are certificates, particularly since we use mutual authentication. Even though we use ECC-based certificates their size can be rather large depending on the content of the certificate and the hierarchy used by the PKI. This aspect has already been recognized in the standardization community and several extensions have been developed or are in development. Unfortunately, those extensions have not found their way into embedded TLS/DTLS stacks.

\begin{figure}
\resizebox{\columnwidth}{!}{
\begin{tikzpicture}
 \pie [polar, explode=0.1, text=legend]
    {43/TLS,
     42/Symmetric Crypto,
     8/HKDF,
     7/RNG}
\end{tikzpicture}
}
\caption{Relative sizes of Flash memory per component, for TLS 1.3 PSK with AES-128-CCM.}
\label{pie:flash:PSK}
\end{figure}
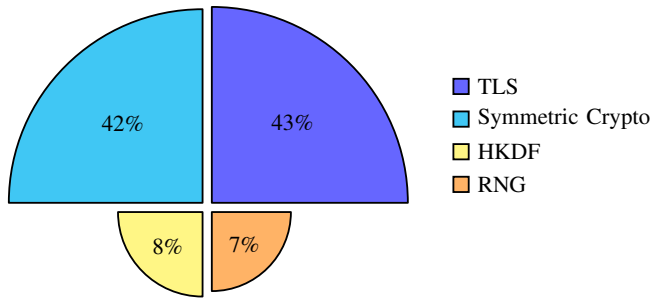

\section{Discussion \& Future Work}
\label{sec:future}

Due to its flexibility, various credential types and extensions can be used with TLS and DTLS. We focused our investigations on PSKs and certificate-based authentication. Quite naturally, PSK-based authentication is superior to certificate-based authentication from a performance point of view. To our surprise, the energy measurements indicate a dramatically lower energy consumption than certificate-based authentication. Particularly when combining DTLS with the newly developed connection ID feature an existing DTLS session can be kept alive for a long time and the need for repeated handshakes can be avoided. 

Our measurements did not take advantage of hardware security features available in many microcontrollers available on the market today. When using these hardware accellerators a significantly lower flash size, RAM utilization, and better energy efficiency can be excepted. It is particularly surprising that the crypto-related software components constitute a significant portion of the overall code size. For a certificate-based TLS configuration, the size of the TLS handshake-related code is small.   

While embedded TLS/DTLS stacks are on the market for a long time already, we can still observe further optimization possibilities on the code level, and various extensions are developed for TLS/DTLS to reduce other parts of the handshake, particularly certificate-related compression schemes (X.509 certificates are still a major contributor to the over-the-air transmission overhead and to increasing heap utilization). 

Thus, our observation is that optimal performance of DTLS and TLS 1.3 on microcontrollers is still to be determined.
And consequently, comparing this performance to that of other solutions (e.g. based on OSCORE and EDHOC) is future work.

On the one hand, EDHOC specification is also work-in-progress. Maturation of this solution is expected to happen next within the newly formed IETF working group LAKE~\cite{LAKE-wg}.
On the other hand hand, work aiming to minimize TLS/DTLS 1.3 footprint is currently in early stages of development in the TLS working group and needs futher evaluation down the line, in order to gauge its full potential.
Recent work has started to focus on minimizing communication security footprint.
For instance, cTLS (Compact TLS 1.3~\cite{I-D.draft-rescorla-tls-ctls})
proposes a configuration of TLS 1.3 which sacrifices backwards compability
for significant savings in over-the-wire size.
This effort leverages both aggressive use of a more compact encoding, and the use of default values for TLS configurations.

\section{Conclusion}
\label{sec:conclusion}

Communication security is a key part of securing IoT devices.
While TLS and DTLS 1.2 protocols have become typical solutions in this context,
TLS 1.3 has recently been published and DTLS 1.3 publication is imminent.

Compared to 1.2, version 1.3 of the protocols provides a number of improvements, including fewer preliminary roundtrips, better privacy protection, and a more modern crypto design. 

In this paper, we have provided an overview of TLS / DTLS 1.3 compared to TLS / DTLS 1.2, via a detailed protocol study comparing many options. 
We also provide a quick overview of on-going related efforts towards minimizing the footprint of communication security above the transport layer.
As such, this paper provide a useful guide for newcomers in this space.

Next, we have conducted an experimental comparative evaluation pf these protocols with respect to stack and heap size, code size, over-the-wire network traffic overhead, and energy consumption through measurements we carried out on commercially available low-power microcontrollers. 

We show that TLS and DTLS 1.3 improvements are accomplished with only small overhead in Flash memory and RAM requirements, compared to TLS/DTLS 1.2. 

As such this paper provides a useful set of data points for a knowledgeable practitioner in this space, to quickly grasp potential issues and make the appropriate choices in terms of IoT software and hardware platform.


\balance
\bibliographystyle{abbrv}
\bibliography{sigproc}

\begin{thebibliography}{10}

\bibitem{RFC5247}
B.~Aboba et~al.
\newblock {Extensible Authentication Protocol (EAP) Key Management Framework}.
\newblock RFC 5247, 2008.

\bibitem{2015-IoT-Lab}
C.~Adjih et~al.
\newblock {FIT IoT-LAB: A Large Scale Open Experimental IoT Testbed}.
\newblock In {\em IEEE World Forum on Internet of Things (WF-IoT)}, pages
  459--464. IEEE, 2015.

\bibitem{2018-RIOT}
E.~Baccelli et~al.
\newblock {RIOT: An Open Source Operating Lystem for low-end Embedded Devices
  in the IoT}.
\newblock {\em IEEE Internet of Things Journal}, 5(6):4428--4440, 2018.

\bibitem{Mint-DTLS}
R.~Barnes.
\newblock {Mint - Minimal (D)TLS Implementation in Go}.
\newblock https://github.com/bifurcation/mint.

\bibitem{baremetal}
H.~Becker.
\newblock {Mbed TLS - Baremetal Branch}.
\newblock https://github.com/ARMmbed/mbedtls/tree/baremetal.

\bibitem{RFC7228}
C.~Bormann, M.~Ersue, and A.~Keranen.
\newblock Terminology for constrained-node networks.
\newblock RFC 7228, 2014.

\bibitem{bossi2016tls13}
S.~Bossi et~al.
\newblock {On TLS 1.3: Early Performance Analysis in the IoT Field}.
\newblock {\em ICISSP}, 2016.

\bibitem{I-D.cragie-tls-ecjpake}
R.~Cragie and F.~Hao.
\newblock Elliptic curve j-pake cipher suites for transport layer security
  (tls).
\newblock Internet-Draft draft-cragie-tls-ecjpake-01, IETF Secretariat, June
  2016.
\newblock
  \url{http://www.ietf.org/internet-drafts/draft-cragie-tls-ecjpake-01.txt}.

\bibitem{ENISA}
ENISA.
\newblock Baseline security recommendations for iot.
\newblock Technical report, November 2017.
\newblock
  \url{https://www.enisa.europa.eu/publications/baseline-security-recommendations-for-iot}.

\bibitem{I-D.friel-tls-atls}
O.~Friel et~al.
\newblock {Application-Layer TLS}.
\newblock Internet-Draft draft-friel-tls-atls-03, IETF Secretariat, July 2019.

\bibitem{hahm2016surveyOS}
O.~Hahm et~al.
\newblock {Operating Systems for Low-End Devices in the Internet of Things: a
  Survey}.
\newblock {\em IEEE Internet of Things Journal}, 2016.

\bibitem{LAKE-wg}
IETF.
\newblock {Lightweight Authenticated Key Exchange (LAKE) Working Group}.
\newblock https://datatracker.ietf.org/wg/lake/about/.

\bibitem{RFC5869}
H.~Krawczyk and P.~Eronen.
\newblock Hmac-based extract-and-expand key derivation function (hkdf).
\newblock RFC 5869, RFC Editor, May 2010.
\newblock \url{http://www.rfc-editor.org/rfc/rfc5869.txt}.

\bibitem{NIST}
NIST.
\newblock {Core Cybersecurity Feature Baseline for Securable IoT Devices: A
  Starting Point for IoT Device Manufacturers}.
\newblock Nistir 8259 (draft), July 2019.
\newblock \url{https://csrc.nist.gov/publications/detail/nistir/8259/draft}.

\bibitem{MbedTLS2.13}
M.~Pégourié-Gonnard.
\newblock {Mbed TLS over low-bandwidth, unreliable datagram networks}.
\newblock https://tls.mbed.org/kb/how-to/controlling\_package\_size.

\bibitem{NSS}
E.~Rescorla.
\newblock {Mozilla NSS DTLS 1.3 Prototype}.
\newblock https://github.com/hannestschofenig/mbedtls/tree/tls13-prototype.

\bibitem{RFC8446}
E.~Rescorla.
\newblock The transport layer security (tls) protocol version 1.3.
\newblock RFC 8446, RFC Editor, August 2018.

\bibitem{I-D.draft-rescorla-tls-ctls}
E.~Rescorla.
\newblock Compact tls 1.3.
\newblock Internet-Draft draft-rescorla-tls-ctls-01, IETF Secretariat, March
  2019.
\newblock
  \url{http://www.ietf.org/internet-drafts/draft-rescorla-tls-ctls-01.txt}.

\bibitem{RFC6347}
E.~Rescorla and N.~Modadugu.
\newblock Datagram transport layer security version 1.2.
\newblock RFC 6347, RFC Editor, January 2012.
\newblock \url{http://www.rfc-editor.org/rfc/rfc6347.txt}.

\bibitem{I-D.ietf-tls-dtls-connection-id}
E.~Rescorla, H.~Tschofenig, and T.~Fossati.
\newblock Connection identifiers for dtls 1.2.
\newblock Internet-Draft draft-ietf-tls-dtls-connection-id-06, IETF
  Secretariat, July 2019.
\newblock
  \url{http://www.ietf.org/internet-drafts/draft-ietf-tls-dtls-connection-id-06.txt}.

\bibitem{Gabriele-RIOT-branch}
G.~Restuccia.
\newblock {Low-power ATLS Test Framework}.
\newblock https://github.com/gabrielication/RIOT/tree/coap\_atls\_adtls.

\bibitem{RFC7924}
S.~Santesson and H.~Tschofenig.
\newblock Transport layer security (tls) cached information extension.
\newblock RFC 7924, RFC Editor, July 2016.

\bibitem{RFC8152}
J.~Schaad.
\newblock Cbor object signing and encryption (cose).
\newblock RFC 8152, RFC Editor, July 2017.

\bibitem{Schneier}
B.~Schneier.
\newblock Security and privacy guidelines for the internet of things.
\newblock Technical report, February 2017.
\newblock
  \url{https://www.schneier.com/blog/archives/2017/02/security_and_pr.html}.

\bibitem{RFC8613}
G.~Selander et~al.
\newblock {Object Security for Constrained RESTful Environments (OSCORE)}.
\newblock RFC 8613, 2019.

\bibitem{RFC7457}
Y.~Sheffer, R.~Holz, and P.~Saint-Andre.
\newblock Summarizing known attacks on transport layer security (tls) and
  datagram tls (dtls).
\newblock RFC 7457, RFC Editor, February 2015.
\newblock \url{http://www.rfc-editor.org/rfc/rfc7457.txt}.

\bibitem{RFC5054}
D.~Taylor, T.~Wu, N.~Mavrogiannopoulos, and T.~Perrin.
\newblock Using the secure remote password (srp) protocol for tls
  authentication.
\newblock RFC 5054, RFC Editor, November 2007.

\bibitem{Securemark}
P.~Torelli.
\newblock {EEMBC SecureMark}.
\newblock https://www.eembc.org/securemark/.

\bibitem{EEMBC}
P.~Torelli.
\newblock {Embedded Microprocessor Benchmark Consortium (EEMBC)}.
\newblock https://www.eembc.org.

\bibitem{Hannes-DTLS13}
H.~Tschofenig.
\newblock {Mbed (D)TLS 1.3 Prototype}.
\newblock https://github.com/hannestschofenig/mbedtls/tree/tls13-prototype.

\bibitem{RFC7925}
H.~Tschofenig and T.~Fossati.
\newblock Transport layer security (tls) / datagram transport layer security
  (dtls) profiles for the internet of things.
\newblock RFC 7925, RFC Editor, July 2016.

\bibitem{RFC7250}
P.~Wouters, H.~Tschofenig, J.~Gilmore, S.~Weiler, and T.~Kivinen.
\newblock Using raw public keys in transport layer security (tls) and datagram
  transport layer security (dtls).
\newblock RFC 7250, RFC Editor, June 2014.

\bibitem{DBLP:journals/access/ZandbergSPTB19}
K.~Zandberg et~al.
\newblock {Secure Firmware Updates for Constrained IoT Devices Using Open
  Standards: A Reality Check}.
\newblock {\em {IEEE} Access, 2019}.

\end{thebibliography}

\end{document}